\documentclass[12pt, twocolumn, numberedappendix, appendixfloats]{openjournal}
\usepackage{lipsum}

\usepackage{xcolor}
\usepackage{textgreek}
\usepackage[utf8]{inputenc}
\usepackage[english]{babel}
\usepackage{amsmath}
\usepackage{hyperref}
\hypersetup{
    unicode, 
    colorlinks=true,
    linkcolor=linkcolor,
    citecolor=linkcolor,
    filecolor=linkcolor,
    urlcolor=linkcolor,
}
\usepackage{color,colortbl}
\definecolor{linkcolor}{rgb}{0.0,0.3,0.5}
\usepackage{tensind}
\tensordelimiter{?}
\DeclareGraphicsExtensions{.bmp,.png,.jpg,.pdf}
\usepackage{verbatim}
\usepackage[normalem]{ulem}
\usepackage{orcidlink}
\usepackage{soul}

\usepackage{CJK}

\usepackage[dvipsnames]{xcolor}
\newcommand{\Msigma}{$M$\text{--}$\sigma$ }

\begin{document}

\title[Cluster Mass Inference from Galaxy Kinematics]{Cluster Mass Inference from Galaxy Kinematics}

\author{Bonny Y.~Wang ({\begin{CJK*}{UTF8}{bsmi}汪玥\end{CJK*}}) \orcidlink{0000-0001-7168-8517}}
\affiliation{Department of Astronomy and Astrophysics, University of Chicago, 5801 S. Ellis Ave., Chicago, IL 60637}
\affiliation{Carnegie Mellon University, 5000 Forbes Ave, Pittsburgh, PA 15213, USA}
\affiliation{Center for Data-Driven Discovery, Kavli IPMU (WPI), UTIAS, The University of Tokyo, Kashiwa, Chiba 277-8583, Japan}
\affiliation{Kavli IPMU (WPI), UTIAS, The University of Tokyo, 5-1-5 Kashiwanoha, Kashiwa, Chiba 277-8583, Japan}

\author{Leander Thiele \orcidlink{0000-0003-2911-9163}}
\affiliation{Center for Data-Driven Discovery, Kavli IPMU (WPI), UTIAS, The University of Tokyo, Kashiwa, Chiba 277-8583, Japan}
\affiliation{Kavli IPMU (WPI), UTIAS, The University of Tokyo, 5-1-5 Kashiwanoha, Kashiwa, Chiba 277-8583, Japan}

\author{Matthew Ho \orcidlink{0000-0003-3207-8868}}
\affiliation{Columbia Astrophysics Laboratory, Columbia University, 550 West 120th Street, New York, NY 10027, USA}


\label{firstpage}

\begin{abstract}
The masses of galaxy clusters carry cosmological and astrophysical information.
We develop a simulation-based inference pipeline to infer cluster masses from full projected phase-space information of member and interloper galaxies.
Our method combines a permutation-invariant Deep Sets architecture with neural posterior estimation using normalizing flows, enabling the recovery of expressive posterior distributions.
We train the model to predict residual corrections to the classical \Msigma relation, thus explicitly isolating information beyond velocity dispersion.
Using the Uchuu–UniverseMachine simulation, we evaluate the method under both idealized (interloper-free) and realistic (cylindrical) observational setups.
In the idealized case, our model reduces the scatter in mass estimates to as low as $\sim 0.1$ dex, representing a twofold improvement over the traditional \Msigma relation.
In the cylindrical setup, we achieve comparable performance at the high-mass end ($> 10^{14.5}\,M_\odot/h$), demonstrating robustness against interloper contamination.
We demonstrate that set-based simulation-driven inference provides a powerful and flexible framework for galaxy cluster mass estimation, enabling improved accuracy and reliable uncertainty characterization for upcoming large-scale surveys.
Our model saturates the kinematic information content and thus suggests a baseline for future studies.
\end{abstract}

\maketitle

\section{\label{sec:level1}Introduction}
Galaxy clusters serve as valuable cosmological tools for constraining key parameters of the Universe \citep{2001ApJ...553..545H, Holder2001,Weller2002,Majumdar2003,2011ARA&A..49..409A,Burenin2012,Mantz2015,Cataneo2015,2016arXiv160407626D, 2020Abbott,Ghirardini2024}. In particular, the abundance and mass distribution of galaxy clusters are sensitive to the growth of large-scale structure, enabling constraints on parameters such as the amplitude of matter fluctuations and the summed mass of neutrinos \citep{Mantz2010,Costanzi2013,Burenin2013, Hou2014, Kirby2019,Fumagalli2024, Bocquet2025}. In light of the current $S_8$ tension between cosmic shear and the cosmic microwave background, precise measurements of the cluster mass function can provide a complementary probe of the standard $\Lambda$CDM framework.

\begin{figure}
    \centering
    \includegraphics[width=0.99\linewidth]{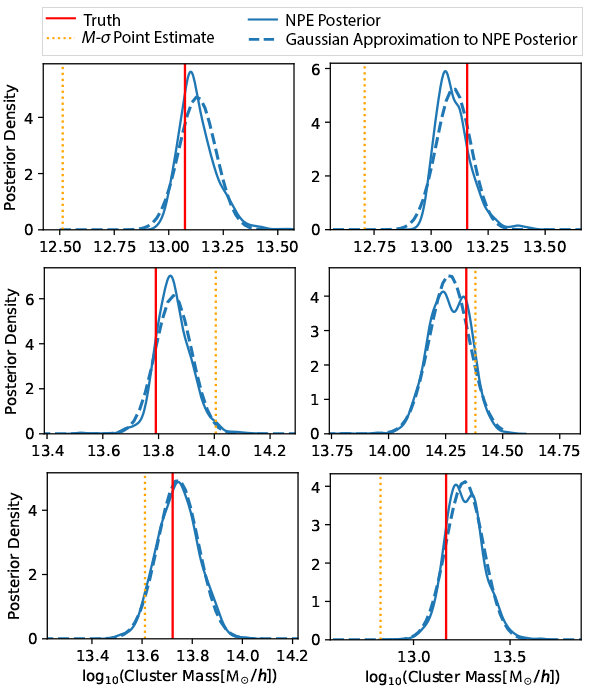}
    \caption{Posterior probability distributions for galaxy cluster masses. Solid blue curves represent the outputs of the neural posterior estimator (NPE) from this work, while dashed blue lines indicate Gaussian fits to the inferred posteriors. The true cluster masses are shown as red vertical lines, and the dotted orange lines denote estimates based on the classical \Msigma relation. The left column displays representative unimodal cases, whereas the right column illustrates examples with bimodal structure.}
    \label{fig:dists}
\end{figure}
A wide range of observational strategies are employed to estimate galaxy cluster masses, focusing on different physical components of clusters. Indirect methods include X-ray observations of the intracluster medium (ICM), which infer cluster masses under assumptions of hydrostatic equilibrium from gas density and temperature profiles \citep[e.g.,][]{Briel1992, Vikhlinin2009, Mantz2016, Giles2017}. Measurements of the thermal Sunyaev-Zel'dovich (tSZ) effect \citep{Sunyaez1970, Sunyaez1972} trace the integrated electron pressure along the line of sight and provide a relatively redshift-independent mass proxy \citep[e.g.,][]{2021ApJS..253....3H}. Optical richness-based estimators, which correlate the number of member galaxies with halo mass, offer an efficient alternative for large surveys but depend sensitively on calibration and selection effects \citep[e.g.,][]{2003ApJ...585..215Y}.

More direct probes of the underlying gravitational potential include weak gravitational lensing, which measures the coherent distortion of background galaxies and is sensitive to the total projected mass distribution independent of cluster dynamical state \citep[e.g.,][]{2014MNRAS.439...48A,2019MNRAS.482.1352M}, as well as galaxy kinematics, which infer masses from the velocity dispersion and phase-space distribution of cluster members \citep[e.g.,][]{1933AcHPh...6..110Z,2014MNRAS.441.1513O,Geller_2013, Shi2024}. Despite their conceptual robustness, these methods face practical limitations: weak lensing measurements often suffer from low signal-to-noise ratios for individual clusters and are affected by projection effects from large-scale structure, while kinematic methods are sensitive to interlopers and deviation from dynamical equilibrium. More broadly, all approaches remain impacted either directly or indirectly by baryonic physics and associated modeling uncertainties, motivating continued efforts to develop complementary methods for robust mass estimation techniques.

In this work, we focus on galaxy cluster dynamics as a complementary and physically motivated avenue for mass estimation, aiming both to improve inference accuracy and to deepen our understanding of the fundamental dynamical processes governing cluster assembly. The classic \Msigma relation provides a simple dynamical mass approximation based on the velocity dispersion of member galaxies. However, it relies on idealized assumptions that clusters are spherically symmetric, composed of equal-mass tracers, and in virial equilibrium. In practice, galaxy clusters exhibit substantial substructure, anisotropies, and ongoing accretion. Thus, they are often far from equilibrium, leading to biased mass estimates from the baseline \Msigma relation. To overcome these limitations, recent studies implement deep learning techniques to model the nonlinear and high-dimensional structure of galaxy phase space \citep[e.g.,][]{Ntampaka2016, Ntampaka2019, Ho2019, Ho2021, Ho2022, Ho2023, Yan2020, KodiRamanah2021, Gupta2020, Gupta2021, Garuda2024, Hanhn2024, Garuda2026, Tominaga2026}. These approaches demonstrate that machine learning (ML) can extract information beyond traditional summary statistics such as velocity dispersion. Nevertheless, many existing methods either compress the data into low-dimensional features, neglect spatial or morphological information, or provide only point estimates without fully characterizing the posterior uncertainty of cluster masses. 

This motivates the development of inference frameworks that preserve richer dynamical information while delivering probabilistic mass estimates. In this work, we adopt the DeepSets architecture~\citep[see e.g.,][]{zaheerDeepSets2017,Thiele2022, Wang2023, Jung2023, DeSanti2023} which naturally accommodates unordered collections of varying sizes and is therefore well suited to the analysis of galaxy cluster member populations. The full projected phase-space information of member galaxies, comprising both line-of-sight velocities and projected positions, is used as input, supplemented by global cluster morphological descriptors. 

The inference framework follows the simulation-based inference (SBI) paradigm \citep{Cranmer2020,Thiele2026}, with the posterior over cluster mass estimated via neural posterior estimation (NPE) \citep{Greenberg2019, Lueckmann2027, Papamakarios2016}. Specifically, neural spline flows \citep{Durkan2019, Dolatabadi2020} are employed as the density estimator, enabling expressive approximations to posterior distributions as shown in Figure \ref{fig:dists}. The predicted posteriors are closer to the true value of cluster masses than those obtained from the classical \Msigma scaling relation. To incorporate physical prior knowledge and improve model interpretability, we train the network to predict residual corrections to the \Msigma relation rather than the cluster mass directly. This procedure explicitly isolates the additional information content beyond that encoded in equilibrium-based assumptions.

This paper is organized as follows. In Section~\ref{sec:method}, we describe the simulation dataset and the simulation-based inference framework, including the construction of the baseline \Msigma\ relation and the architecture of the machine learning pipeline. In Section~\ref{sec:idealized}, we present the idealized setup, in which only true cluster members are considered, and evaluate the performance of the model in this controlled setting. In Section~\ref{sec:cylinder}, we extend the analysis to a more realistic cylindrical setup that includes interloper galaxies, and assess both interloper identification and mass inference performance. Finally, we summarize our main findings and discuss implications in Section~\ref{sec:conclusion}.

\section{Method}
In this section, we describe the simulation dataset and the simulation-based inference framework used for galaxy cluster mass estimation. We first introduce the Uchuu--UniverseMachine dataset and establish the baseline \Msigma\ relation as a reference model. We then detail the machine learning architecture, including the Deep Sets feature extractor and the normalizing flow used for posterior estimation, as well as the specific model setup for mass inference and interloper identification.

\label{sec:method}
\subsection{Dataset}

\begin{figure}[!htb]
    \centering
    \includegraphics[width=0.8\linewidth,trim={2cm 1cm 1.5cm 0},clip]{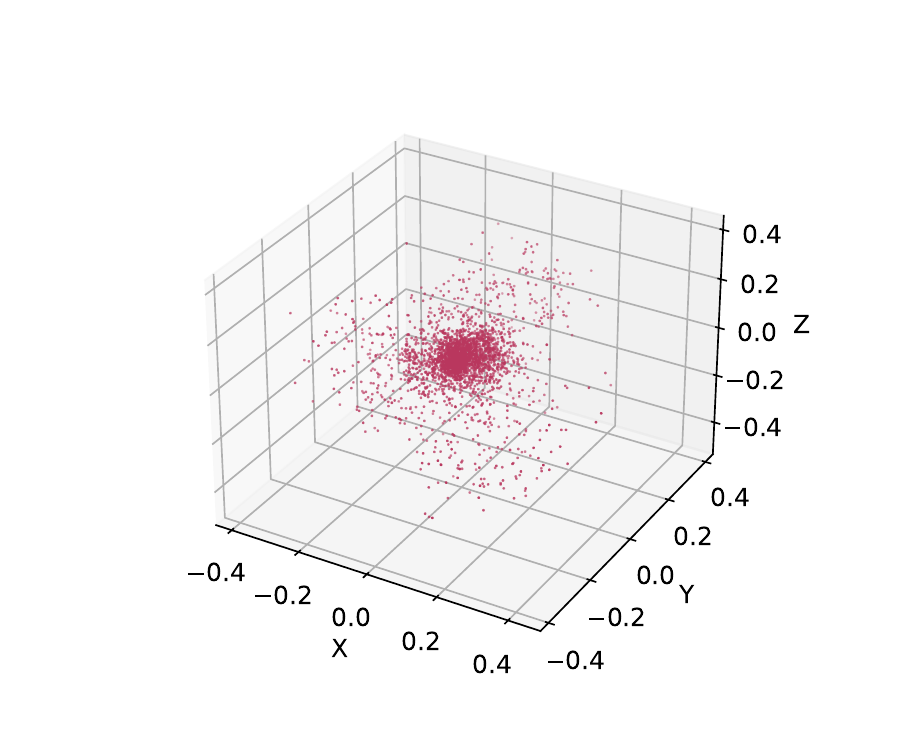}
  
    \caption{Sample cluster from the Uchuu-UniverseMachine datase. Each magenta point represents a galaxy inside the cluster.}
    
    \label{fig:cluster-sample}
\end{figure}
To ensure an accurate representation of galaxy properties and account for the large scale of galaxy clusters, we select the Uchuu-UniverseMachine dataset \citep{Ishiyama2021,Aung2023} at redshift 0.5 for this study. Uchuu is one of the largest and most detailed cosmological simulations available. It encompasses a comoving volume of approximately $(2\, \text{Gpc}/h)^3$, resolved with $12,800^3$ dark matter particles. For our analysis, we utilize host halos from the Rockstar halo catalog \citep{Behroozi2013}, combined with galaxy properties from the UniverseMachine model \citep{Behroozi2019}.

Figure~\ref{fig:cluster-sample} illustrates a sample cluster from our dataset, showing galaxies as magnenta dots in normalized coordinates. 

\begin{figure}
    \centering
    \includegraphics[width=0.99\linewidth]{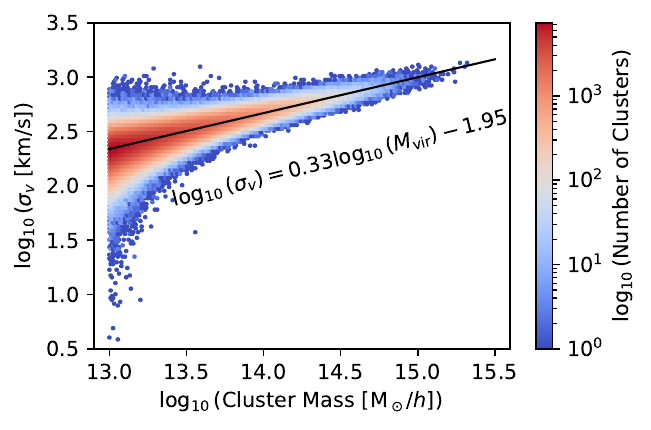}
    \caption{Relationship between the virial mass $M_{\text{vir}}$ and the random line-of-sight velocity dispersion $\sigma_v$ of member galaxies in our dataset (idealized setup). The color scale represents the number density of clusters in each bin. The data approximately follow a linear trend, and the best-fit regression line is given by $\log_{10}(\sigma_v) = 0.33 \log_{10}(M_{\text{vir}}) - 1.95$.}
    \label{fig:M_sigma}
\end{figure}

In Figure~\ref{fig:M_sigma}, we present the relationship between the virial mass ($M_{\text{vir}}$) and the random line-of-sight velocity dispersion ($\sigma_v$) of member galaxies identified by the Rockstar halo finder in our clusters. This demonstrates the traditional \Msigma relation. The data follow an approximately linear trend, expressed as 
\begin{equation}
\label{kb_eq}
\log_{10}(\sigma_v) = 0.33 \log_{10}(M_{\text{vir}}) - 1.95 + \gamma.
\end{equation}
where $\gamma$ is a random variable drawn from a Gaussian distribution with a fixed variance and we determined the parameters using linear regression.
This serves as our baseline for understanding the connection between galaxy dynamics and cluster mass.
We note that the specific values of slope and intercept can vary depending on the fitted mass range, sampling of clusters, and the choice of regression objective.
However, we have verified that reasonable perturbations to the parameters do not qualitatively affect our downstream results or conclusions.

\begin{figure}
    \centering
    \includegraphics[width=0.9\linewidth]{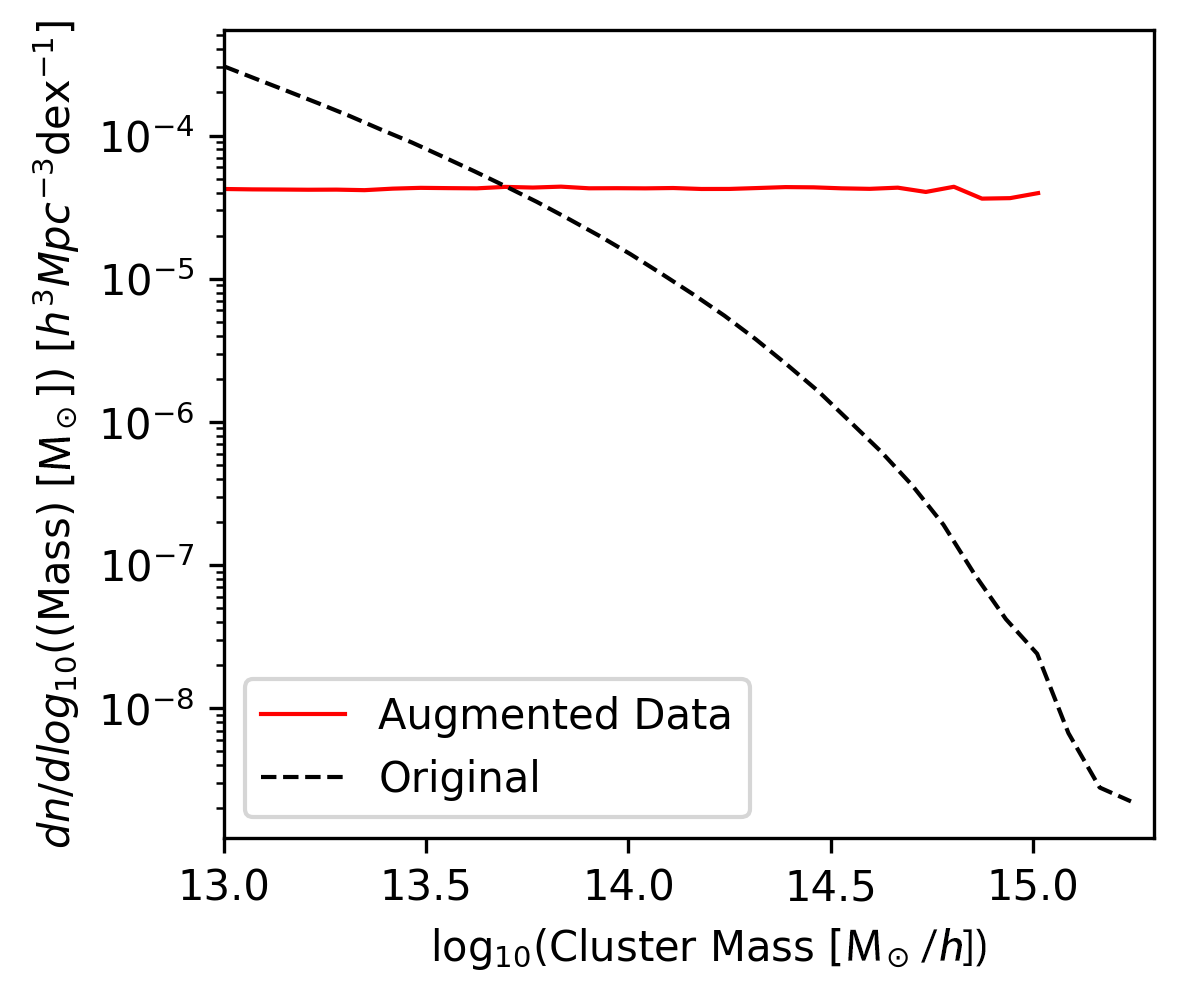}
    \caption{Comparison of cluster mass functions. The black dashed line shows the original distribution from the simulation, while the red solid line represents the augmented data, which are uniformly distributed.}
    \label{fig:mass}
\end{figure}

In addition, to mitigate sampling biases introduced by the halo mass function and ensure balanced representation across all mass scales, we applied data augmentation by randomly selecting different directions of the line-of-sight to construct a training set with approximately uniform mass distribution. This is shown in Figure \ref{fig:mass}.

\begin{figure*}[!htb]
    \centering
    \includegraphics[width=0.999\linewidth]{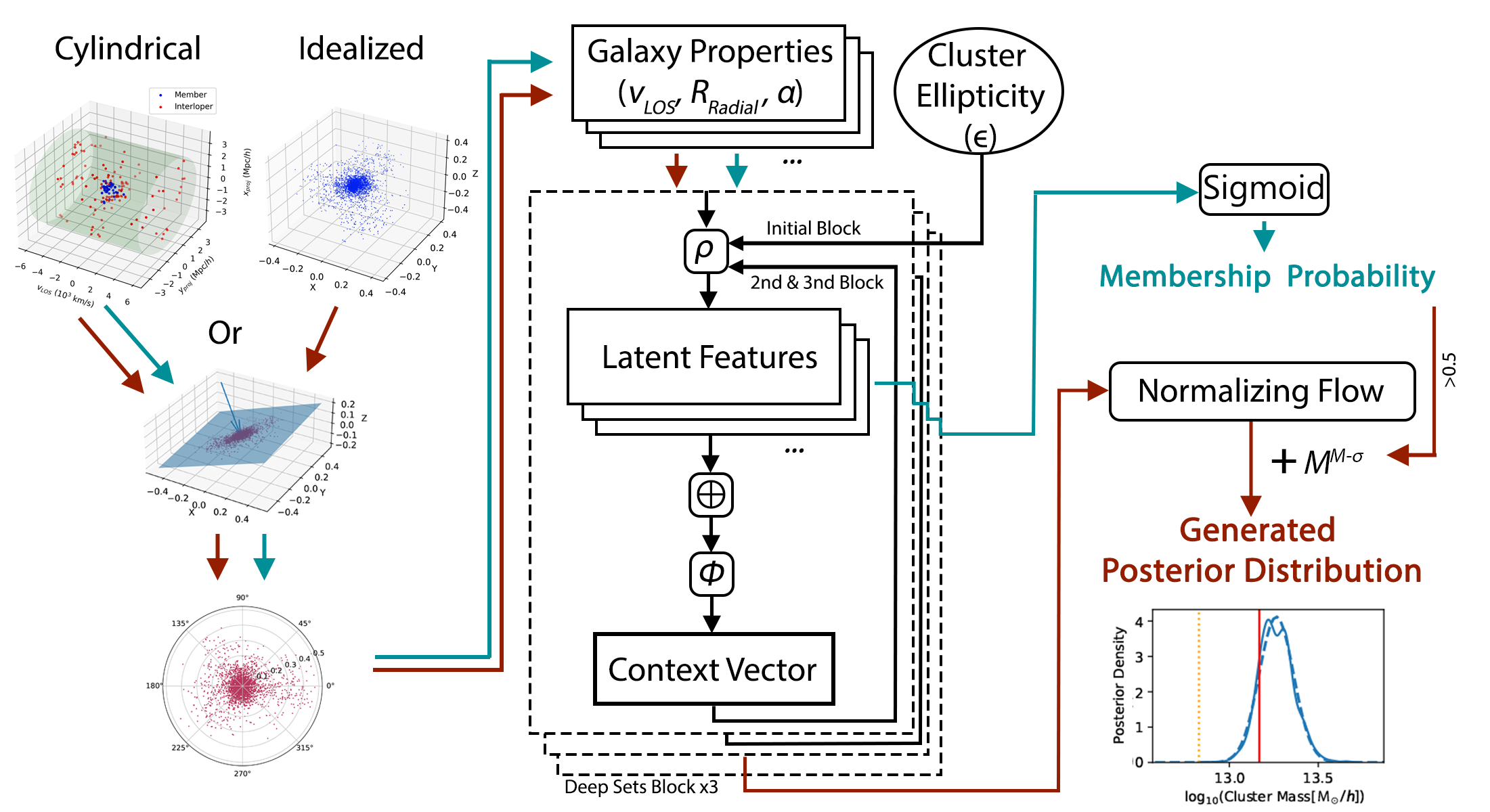}
    \caption{Schematic overview of the model for interloper identification and probabilistic cluster mass estimation. Galaxy observables---line-of-sight velocity $v_{\mathrm{LOS}}$, projected radius $R_{\mathrm{radial}}$, and angular position $\alpha$---are constructed by projecting galaxies onto randomly oriented planes for both idealized and cylindrical catalogs. For mass inference (red arrows), these features are processed through three Deep Sets blocks to produce permutation-invariant latent representations. The first block incorporates the global cluster property, ellipticity $\epsilon$, which is propagated and updated through subsequent layers to form a final context vector. This vector conditions a normalizing flow, yielding the full posterior distribution of the cluster mass as a residual correction to the $M\text{--}\sigma$ relation. For cylindrical catalogs, an additional membership identification network (cyan arrows) that shares the same Deep Sets architecture is applied prior to mass inference. This network outputs galaxy membership probabilities via a sigmoid function. Galaxies with membership probability $>0.5$ are then used to construct the baseline \textbf{$M\text{--}\sigma$} mass estimate.}
    \label{fig:schematic}
\end{figure*}

\subsection{ML Setup}
\label{sec:ML_setup}
To assess how much additional information can be extracted from galaxy dynamics beyond the baseline \Msigma relationship, we train a machine learning model to predict the posterior density over the residual between the true virial mass and the baseline prediction. The corrected virial mass is then obtained by applying this machine learning adjustment to the original \Msigma estimate. We express this as:

\begin{equation}
    \log_{10} M_{\text{vir}}^{\text{pred}} = \log_{10} M_{\text{vir}}^{M\text{-}\sigma} + \Delta_{\text{ML}},
\end{equation}
where \( M_{\text{vir}}^{M\text{-}\sigma} \) is the mass estimated from the traditional \Msigma relation using equation \ref{kb_eq}, and \( \Delta_{\text{ML}} \) is the machine learning correction term. This setup allow us to explicitly isolate information beyond what is captured by equilibrium-based assumptions.

Our machine learning model comprises two main components: a deep sets model for extracting information from the galaxies in the cluster on which a normalizing flow model is conditioned to generate full posterior distributions. 

Each Deep Sets block consists of a pointwise encoder \( \rho \), and a global context updater \( \phi \), both modeled with multilayer perceptrons (MLPs). A permutation-invariant aggregation operation, denoted \( \bigoplus \), ensures that the model respects the unordered nature of the input set. We use mean pooling for aggregation to avoid encoding cluster richness as an implicit feature. We can express our model as follows. 

Let \( \vec{p}_i \) represent the feature vector of the \( i \)-th galaxy, and \( \vec{u}^{(0)} \) denote the initial cluster-level context vector. The operations within the \( k \)-th Deep Sets block are defined as:
\begin{align}
    \label{eq:Deepset}
    \vec{h}_i^{(k)} &= \rho^{(k)}\left([\vec{p}_i, \vec{u}^{(k-1)}]\right), \quad 
    \vec{u}^{(k)} = \phi^{(k)}\left( \bigoplus \vec{h}_i^{(k)} \right), 
\end{align}
where \( \vec{h}_i^{(k)} \) is the encoded representation of galaxy \( i \) at level \( k \), and \( \vec{u}^{(k)} \) summarizes the cluster dynamics up to that stage. After passing through three stacked Deep Sets blocks, the final global feature vector \( \vec{u}^{(3)} \) encodes the latent dynamical state of the cluster.

To achieve uncertainty quantification, we extend our architecture with a generative model known as normalizing flows \citep{Jimenez2015}. This model uses neural networks to learn a bijective mapping, \( f : x \mapsto z \), which transforms a complex target distribution into a simpler base distribution, \( \pi(z) \). The mapping \( f \) is designed to be invertible with a tractable Jacobian, enabling the target distribution to be derived from \( \pi(z) \) via a change of variables. In our application, the target distribution corresponds to the posterior of cluster mass, while the base distribution is modeled as a univariate standard Gaussian. We use the final global feature vector \( \vec{u}^{(3)} \) from the deep sets model as a conditioning variable for this generative model. For our specific implementation, we used the conditional spline model in \textsc{Pyro} package~\citep{Bingham2018Pyro}.

By combining these two models, the deep sets for feature extraction and the normalizing flows for posterior generation, we are able to predict the full posterior distribution of galaxy cluster mass from member galaxy dynamics. We jointly train both models by maximizing the log probability of the posterior:

\begin{equation}
    \mathcal{L} = -\log p(\Delta_{\text{ML}} \mid \vec{u}^{(3)}).
\end{equation}

In the case of a contaminated catalog containing interloper galaxies, described in detail in Section~\ref{sec:cylinder}, we train a dedicated classification network to identify interlopers prior to the mass inference stage. We adopt the same hierarchical Deep Sets architecture defined in Equation~\ref{eq:Deepset}, but discard the normalizing flow part. We directly use the individual latent representation $\vec{h}_i^{(3)}$, which encodes both the local properties of that galaxy and its global context within the cluster population through the aggregated context vector $\vec{u}^{(k)}$. This latent vector is passed directly through a sigmoid activation function ($\varsigma$) to produce a scalar membership probability,
\begin{equation}
    \vec{y}_i = \varsigma\!\left(\vec{h}_i^{(3)}\right),
    \label{eq:cls_head}
\end{equation}
where $\vec{y}_i \in [0, 1]$ denotes the predicted probability that galaxy $i$ is a true cluster member. We then use the Binary Cross-Entropy (BCE) Loss for training this classifier. At inference time, galaxies with $\vec{y}_i \geq 0.5$ are retained as cluster members. Those below this threshold are flagged as interlopers. The performance of our interloper identification network is shown in detail in section \ref{sec:cylindircal_predictions} as well as in Appendix \ref{app:interloper_performance}.

Once trained, the interloper classifier is incorporated as a fixed preprocessing stage in the full mass inference pipeline above. Only the identified member properties is used to calculate the $\log_{10} M_{\text{vir}}^{M\text{-}\sigma}$. However, notably, interlopers are still included in the pipeline as they might encode some neighborhood information which could benefit the mass estimation of the cluster. Therefore, $\Delta_{\text{ML}}$ is predicted using information from both interlopers and true members of the cluster.

In the following two sections, we present in detail the input features and training results, both with and without the inclusion of interloper galaxies. We refer to the dataset without interlopers as the \textit{idealized} setup, and the dataset including interlopers as the \textit{cylindrical} setup.

\section{Idealized setup}
We begin by evaluating our framework in an idealized setting, where only true cluster member galaxies are included. This setup provides a controlled environment to isolate the intrinsic performance of the model without contamination from interlopers and any additional observation selection bias. 
\label{sec:idealized}
\subsection{Feature Construction}
\label{sec:Spherical_FC}
In an idealized setup, we consider only the member galaxies of the host cluster (identified by the Rockstar halo finder), without any contamination from interloper galaxies. To mimic the observation process, we use the following process to construct our dataset. First, we select a random line of sight and define a plane perpendicular to it. We then project the galaxies onto this plane. To capture the overall shape of the cluster while ensuring rotational invariance within the plane, we fit an ellipse to the projected galaxy distribution and use its semimajor axis as the reference axis. We show the above steps on a sample cluster in the Idealized part of Figure \ref{fig:schematic}. Following this setup, we select the following node features as inputs for our machine learning model:
    \begin{enumerate}
        \item Line-of-sight velocity, $v_{\rm LOS}$,
        \item Projected radial distance to the center of the halo cluster, $R_\perp$,
        \item Angle between the galaxy position and the fitted axis from the ellipse, $\alpha$.
    \end{enumerate}
In addition, we use a global feature to characterize the overall distribution of galaxies, namely the ellipticity $\epsilon$ calculated from the semi-major axis and the semi-minor axis from the fitted ellipse.

\begin{figure*}[!htb]
    \centering
    \includegraphics[width=0.85\linewidth]{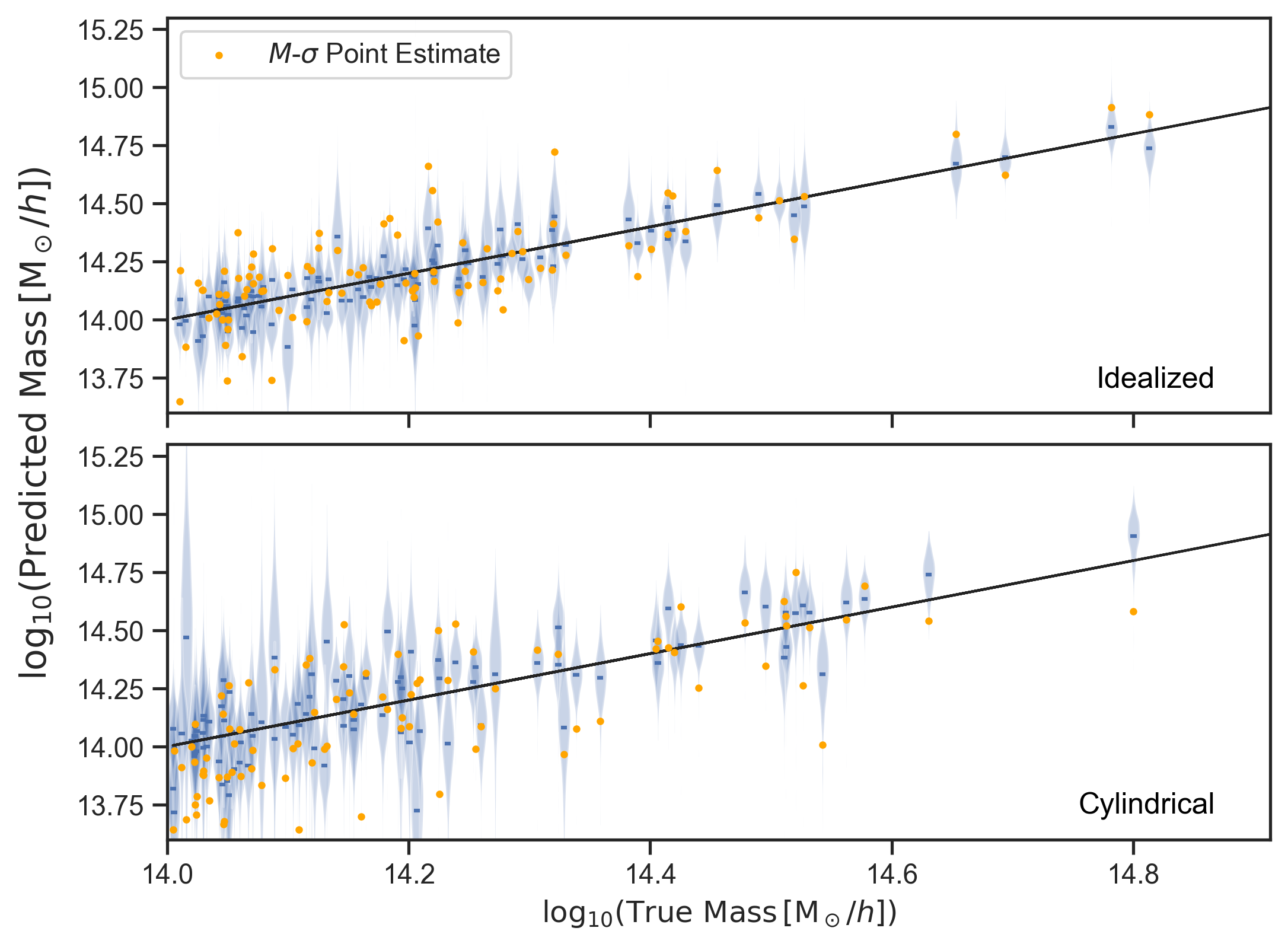}
    \caption{Comparison of predicted and true galaxy cluster masses. The x-axis represents the true mass values and the y-axis shows the predicted masses obtained through our machine learning. The orange points mark the viral prediction from galaxies' velocity dispersion, while the violin plots illustrate the generated posterior distributions. The black reference line shows when true and predicted values perfectly match. The text in the bottom right indicates whether the result is from the idealized or cylindrical setup.}
    \label{fig:violin-sphere}
\end{figure*}
\begin{figure*}[!htb]
    \centering
    \includegraphics[width=0.49\linewidth]{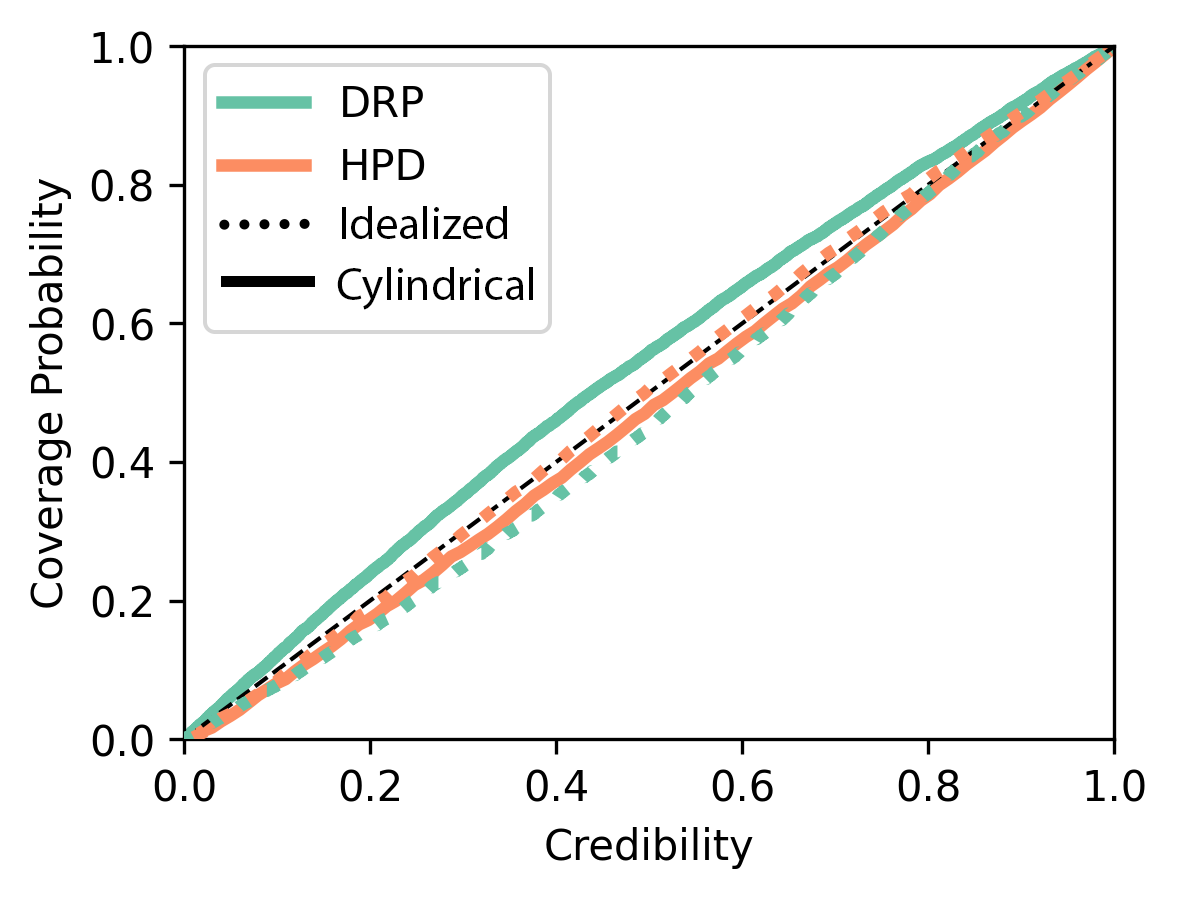}
    \includegraphics[width=0.49\linewidth]{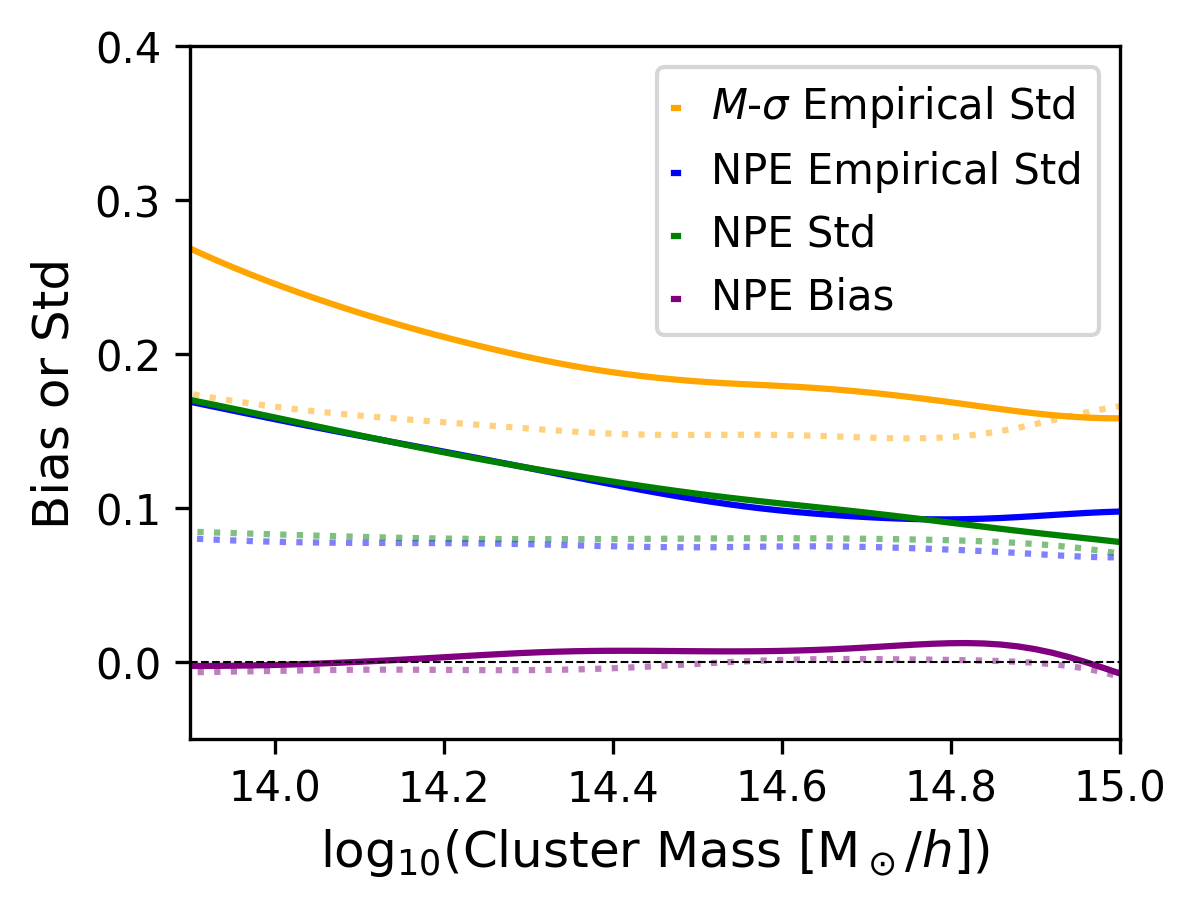}
    \caption{Tests of the trained models' performance. \textbf{Left}: Comparison of coverage probabilities against credibility levels for TARP and HPD methods, where HPD (orange) represents the highest posterior density coverage test and TARP (green) represents the distance to random point method. \textbf{Right}: Bias and scatter in predicted cluster masses as a function of true mass. The orange and blue curves represent the standard deviation of the \Msigma relation and the NPE mean estimates relative to the true masses, respectively. The green curve shows the average predicted posterior standard deviation, while the purple line indicates the mean prediction bias across mass bins. The dashed black line marks the zero-bias reference.}
    \label{fig:performance}
\end{figure*}

\subsection{ML Predictions}
\label{sec:Spherical_ML_Pred}
Using the galaxy-level and global features described above, we train our machine learning model as outlined in the previous section. The top panel of Figure~\ref{fig:violin-sphere} compares the predicted and true cluster masses. The x-axis shows the true masses, while the y-axis shows the predicted masses from our model. Each orange point represents the traditional virial mass estimate derived from the \Msigma relation. The surrounding violin plots visualise the full posterior distributions produced by the model, capturing the predicted uncertainty. The black diagonal line denotes perfect agreement between predicted and true values. Most distributions are centred near this line, indicating that the model's predictions are generally consistent with the true masses and perform comparably to or better than traditional virial estimates.
 
To assess uncertainty calibration, we apply two coverage tests: the traditional HPD (Highest Posterior Density) method and the TARP (Tests of Accuracy with Random Points) method \citep{Lemos2023},  which serves as an alternative to HPD by estimating coverage probabilities from random reference points in parameter space. Figure~\ref{fig:performance} compares the coverage probabilities across credibility levels for both methods, with the spherical setup shown as dotted lines. The close alignment of the HPD (red) and TARP (blue) curves with the diagonal reference line indicates that the predicted posteriors are well-calibrated.
 
The right panel of Figure~\ref{fig:performance} presents a detailed comparison of both the bias and standard deviation in predicted cluster masses relative to their true values, again with dotted lines for the spherical setup. The blue dotted curve shows the empirical scatter of the ML predictions, while the green dotted curve shows the posterior standard deviation predicted by the model. These two curves remain closely aligned across nearly the entire mass range, further confirming that the model provides well-calibrated uncertainty estimates. In contrast, the orange dotted curve, representing the standard deviation of the traditional $M$--$\sigma$ point estimate, exhibits significantly larger scatter. Notably, the empirical standard deviation from NPE is approximately half that of the $M$--$\sigma$ relation, achieving lognormal residuals below 0.1 dex. This reflects a substantial improvement in mass prediction precision. The purple dotted line, tracking the average prediction bias, remains close to zero across all mass bins, demonstrating that the model does not exhibit systematic bias.

 \begin{figure}
    \centering
    \includegraphics[width=0.995\linewidth]{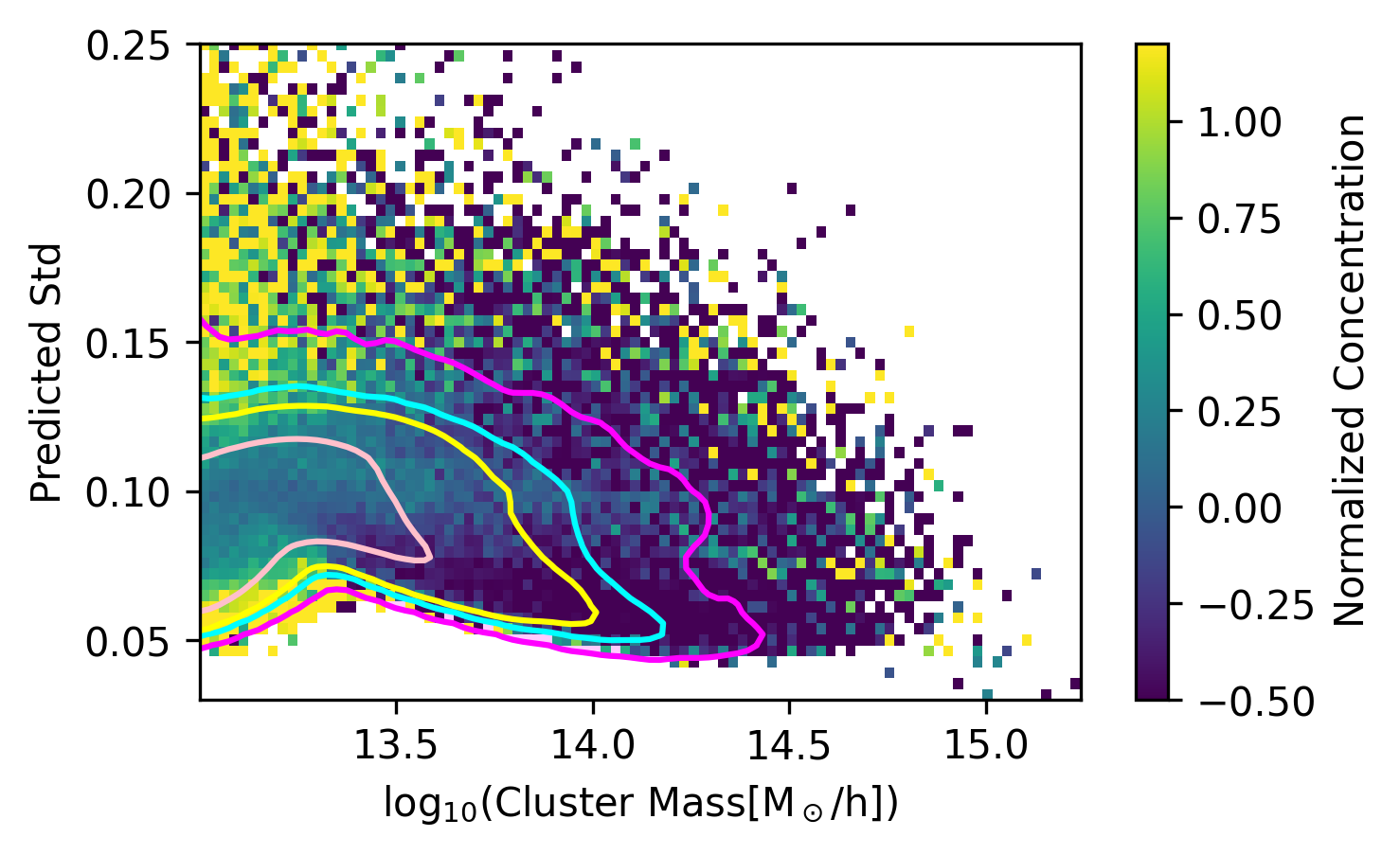}
    \caption{Predicted posterior standard deviation as a function of true cluster mass, color-coded by normalized cluster concentration. The overlaid contours indicate the 68\%, 90\%, 94\%, and 98\% levels of the number density distribution, revealing a bimodal structure: the upper peak corresponds to higher concentrations, while the lower peak corresponds to lower concentrations.}
    \label{fig:concentration}
\end{figure}

\subsection{Interpretation of trained model}
To better understand the internal behavior of our machine learning model, we further investigate how additional physical properties of galaxy clusters influence its predictions. In particular, we focus on the relationship between the predicted posterior uncertainty and the concentration of clusters.
Concentration is known to correlate with the dynamical state of galaxy clusters and thus makes for an interesting secondary proxy to consider \citep{Wechsler2002, Neto2007,Cassano2010, Yuan2020}.

Figure~\ref{fig:concentration} presents the predicted posterior standard deviation as a function of the true cluster mass, with each point color-coded by the cluster's normalized concentration. This normalization is performed to remove the inherent mass dependence of concentration, allowing us to isolate residual trends. To further visualize the distribution of clusters in this two-dimensional space, we overlay contours that represent the number density. These contours reveal a bimodal distribution: one peak appears at higher concentration values (the upper ridge), and another at lower concentrations (the lower ridge). Previous studies have suggested that the fraction of unrelaxed clusters increases with mass \citep{Ludlow2013}, which may explain how the bimodal structure becomes more pronounced at higher cluster masses.

Interestingly, the model appears to provide more precise (i.e., lower uncertainty) mass estimates for clusters with lower concentrations, which are typically interpreted as being less dynamically relaxed. This finding runs counter to the expectation that dynamically relaxed clusters -- often characterized by higher concentrations -- should be easier to model under the assumption of virial equilibrium. One possible explanation is that some highly concentrated clusters may actually be in non-equilibrium states or in a transitional phase of their evolution, possibly due to recent mergers or other dynamical disturbances \citep{Ludlow2012, Sereno2013}.

\section{Cylinder setup}
In addition to the idealized spherical setup, we also evaluate our model on a more realistic mock observational scenario that includes interloper galaxies and perform cuts on galaxie stellar mass. This setup mimics a typical observational setting and presents a more challenging environment for both membership identification and mass inference.

\label{sec:cylinder}
\subsection{Feature Construction}
\label{sec:cylinder_feature_construct}
This dataset is constructed by applying cylindrical cuts around each galaxy cluster, similar to the procedure described in \cite{Ho2019}. There are only two deviations from \cite{Ho2019}. First, rather than padding the simulation box, we apply full periodic boundary corrections to galaxies whose displacement from the target cluster exceeds half the box length, 1000 $\mathrm{Mpc}/h$. We do not expect this difference to have a significant impact on the resulting catalog. Second, when examining the public code of \cite{Ho2019},\footnote{\url{https://github.com/McWilliamsCenter/halo_cnn}} we found that central galaxies of neighboring halos are excluded when creating contaminated catalogs. We add back these galaxies, since they can also act as interlopers and we found no observational properties that would distinctly justify excluding them. This choice increases the interloper fraction by approximately 20\%. The effect may be particularly important at the low-mass end where clusters have lower richness and are therefore more sensitive to additional contaminants. We discuss the impact of this increased contamination on the cylindrical mass predictions in Section~\ref{sec:cylindircal_predictions}. We briefly summarise the rest of the key steps here. 

A galaxy is retained as a cylindrical candidate if it satisfies: 
a projected spatial cut
\begin{equation}
    R_\perp \leq 3.5\;h^{-1}{\rm Mpc},
\end{equation}
where $R_\perp$ is the distance from the galaxy to the 
cluster center projected onto the plane perpendicular to the line-of-sight 
direction $\hat{n}$,
and a stellar mass ($M_*$) threshold
\begin{equation}
    \log_{10}(M_* / h^{-1}M_\odot) \geq 9.5.
\end{equation}

\begin{figure}
    \centering
    \includegraphics[width=0.95\linewidth]{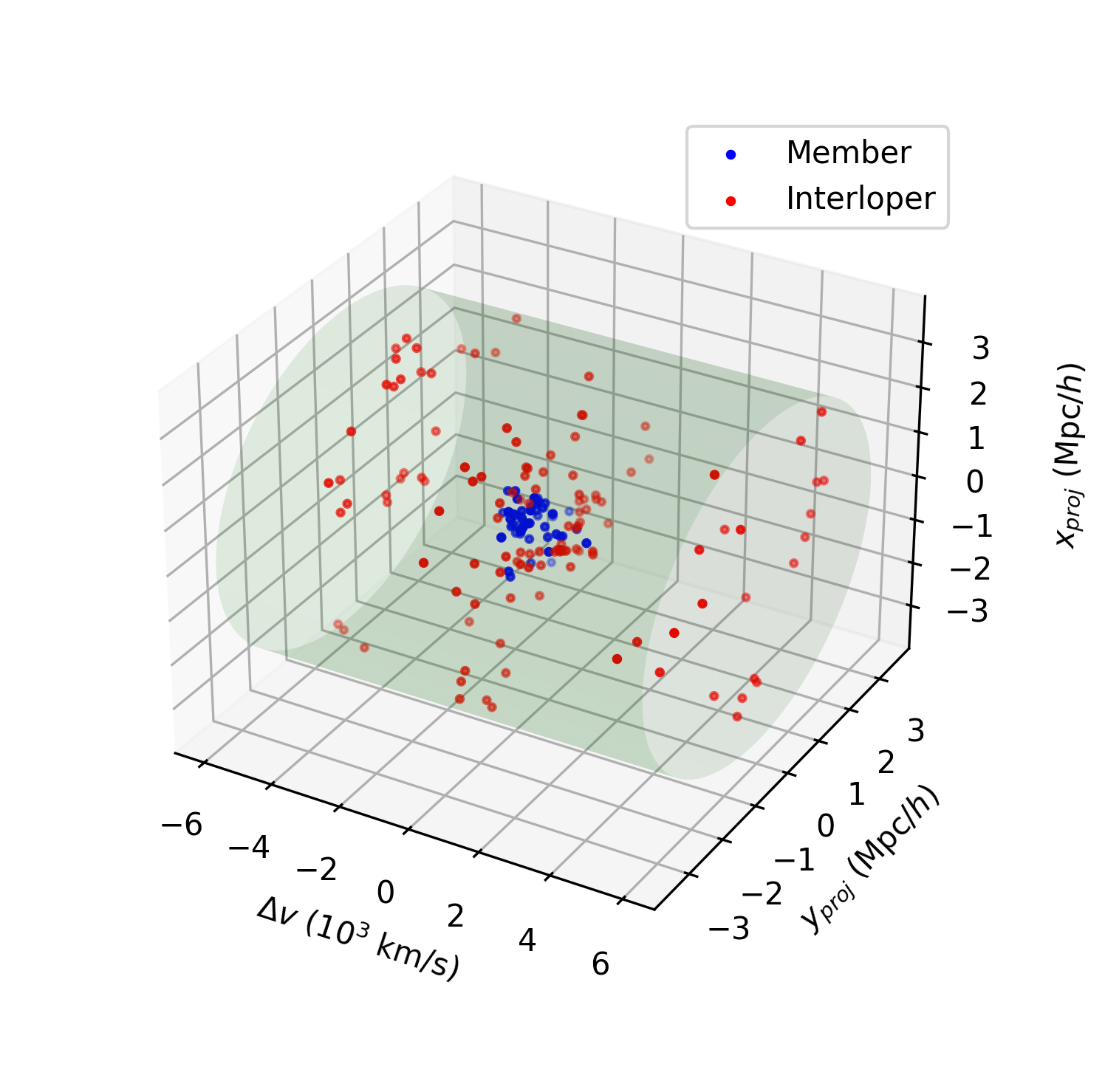}
    \caption{Three-dimensional phase-space distribution of galaxies within a cylindrical aperture around a representative cluster. Blue points indicate true cluster members identified by the Rockstar halo finder, while red points denote interloper galaxies. All galaxies also satisfy $\log_{10}(M_* / h^{-1}M_\odot) \geq 9.5$. The axes show the calculated velocity $\Delta v$ and the two projected spatial coordinates $x_{\rm proj}$ and $y_{\rm proj}$. The green shaded region illustrates the cylindrical selection volume, defined by a projected radial cut of $R_\perp \leq 3.5\,h^{-1}\,\mathrm{Mpc}$ and a velocity cut of $|\Delta v| \leq 6000\,\mathrm{km\,s^{-1}}$.}
    \label{fig:Cylinder-Setup}
\end{figure}

We then recompute the observed line-of-sight velocity of each galaxy relative to its host cluster following the appendix in \cite{Ho2019}, to which we refer the reader for a full description. We briefly summarise the procedure in three steps below. First, the displacement between a galaxy and its host cluster ($\mathbf{r}_{\rm gal} - \mathbf{r}_{\rm clu}$) is computed with periodic 
boundary conditions and projected onto the unit line-of-sight vector $\hat{n}$:
\begin{equation}
    d_{\rm LOS} = (\mathbf{r}_{\rm gal} - \mathbf{r}_{\rm clu}) \cdot \hat{n}.
\end{equation}
The galaxy's comoving distance is then
\begin{equation}
    d_{\rm gal} = d_{\rm clu} + d_{\rm LOS},
\end{equation}
from which the galaxy and cluster redshift $z_{\rm gal}$ are obtained via interpolation. We assume all clusters are located at redshift 0.5 and calculate the $d_{\rm clu}$ based on this redshift. Second, the Hubble recession velocity at redshift $z$ is computed relativistically as
\begin{equation}
    v^{\rm H}(z) = \frac{(1+z)^2 - 1}{(1+z)^2 + 1}\,c,
\end{equation}
giving $v^{\rm H}_{\rm clu} = v^{\rm H}(z_{\rm clu})$ and 
$v^{\rm H}_{\rm gal} = v^{\rm H}(z_{\rm gal})$ for the cluster and galaxy, respectively. Finally, we perform a relativistic velocity combination, combining its peculiar motion and its Hubble recession velocity:
\begin{equation}
    \Delta v = \left(v^{\rm LOS}_{\rm gal} + v^{\rm H}_{\rm gal}\right) 
    - \left(v^{\rm LOS}_{\rm clu} + v^{\rm H}_{\rm clu}\right),
\end{equation}
where $+$ and $-$ denote the relativistic velocity addition and 
subtraction operators. Based on this value, we apply a final velocity cut 
\begin{equation}
    |\Delta v| \leq 6000\;{\rm km\,s}^{-1}
\end{equation}
to yield the final galaxy sample within the cylindrical aperture. In addition, clusters with fewer than 10 members are discarded. We show a visualization of this cylindrical cut in Figure \ref{fig:Cylinder-Setup}.

After this cut, we use the same procedure as described in section \ref{sec:Spherical_FC} to construct the spatial node features (radial distance $R_{radial}$ and angle $\alpha$) and the global feature (ellipticity $\epsilon$). We replace the line-of-sight velocity $v_{LOS}$ with the calculated $\Delta v$. These features are passed to train both the interloper identification network and then the mass inference network as described in section \ref{sec:ML_setup}.

\subsection{ML Predictions}
\label{sec:cylindircal_predictions}

\begin{table*}[tp]
\centering
\caption{Comparison of statistical metrics between the Caustic and ML methods across mass bins.}
\label{tab:method_comparison}
\renewcommand{\arraystretch}{1.3}
\begin{tabular}{l ccccc ccccc}
\hline\hline
& \multicolumn{5}{c}{\textbf{Caustic Method}} & \multicolumn{5}{c}{\textbf{ML Method}} \\

\textbf{Mass Bin} & Acc. & Prec. & Recall & Contam. & $F_1$ & Acc. & Prec. & Recall & Contam. & $F_1$ \\
$[\log_{10}(\mathrm{M}_\odot/h)]$ & & & & & & & & & & \\
\hline
$14.00$--$14.25$ & 0.707 & 0.359 & 0.838 & 0.641 & 0.503 & 0.921 & 0.729 & 0.862 & 0.371 & \textbf{0.797} \\
$14.25$--$14.50$ & 0.651 & 0.403 & 0.853 & 0.597 & 0.548 & 0.885 & 0.729 & 0.866 & 0.371 & \textbf{0.790} \\
$14.50$--$14.75$ & 0.614 & 0.460 & 0.868 & 0.541 & 0.601 & 0.856 & 0.752 & 0.856 & 0.330 & \textbf{0.787} \\
$14.75$--$\infty$ & 0.583 & 0.493 & 0.917 & 0.507 & 0.642 & 0.804 & 0.766 & 0.836 & 0.305 & \textbf{0.772} \\
\hline
\end{tabular}

\end{table*}

Here, we present the performance of both the interloper identification model and the full concatenated pipeline with the cluster mass inference.

In order to assess the performance of our ML interloper identification network, we compare it against the analytical caustic method \citep{Sereno2013}, implemented via \texttt{CausticSNUpy} \citep{Kang2024}, in which interlopers are rejected based on their location relative to the caustic amplitude in projected phase space. The results are summarized in Table~\ref{tab:method_comparison}, which reports accuracy, precision, recall, contamination, and $F_1$ score across four halo mass bins. 

The ML method consistently outperforms the caustic approach in all mass bins, achieving $F_1$ scores in the range 0.772 to 0.797, compared to 0.456 to 0.634 for the caustic method, and maintaining a more balanced set of classification metrics overall. However, we note that the comparatively poorer performance of the caustic method is likely driven by a mismatch between the method and the specific properties of our dataset. The caustic technique is not explicitly tuned for this data configuration and relies on a simplified, effectively one-dimensional phase-space characterization of cluster structure. In contrast, our dataset and model incorporate richer, higher-dimensional information, including full 2D morphological features. As a result, the caustic method may struggle to robustly recover the escape velocity profile under these conditions, particularly in lower-mass systems where the available phase-space sampling very limited (around 40 galaxies for $10^{14} \mathrm{M}_\odot/h$ clusters ).

A more detailed illustration of the caustic method and additional performance analysis are presented in Appendix~\ref{app:interloper_performance}, as the interloper network presented here is used primarily to generate the benchmark \Msigma estimates employed in subsequent analysis.

After training the interloper identification network, we use the identified member galaxies together with the newly computed $\Delta v$ values to recalibrate the baseline $M\text{--}\sigma$ relation for the cylindrical setup, obtaining slope 0.34 and intercept 2.08. This recalibrated relation now serves as the baseline mass estimator, and the machine learning model learns the residual correction $\Delta_{\text{ML}}$ to its predictions.

We assess the mass inference from the cylindrical setup using the same analysis framework described for the spherical case in Section~\ref{sec:Spherical_ML_Pred}. We emphasize here that the interloper identification is only used for generating the baseline \Msigma prediction and the full mass inference pipeline retains both member and interloper information as described in Section~\ref{sec:ML_setup}.

Similar to the idealized setup, the bottom panel of Figure~\ref{fig:violin-sphere} presents the model performance in the presence of interlopers. Compared to the idealized case, the predictions exhibit increased scatter around the one-to-one relation. The posterior distributions are correspondingly broader. Nevertheless, with most posteriors close to and covering the reference line the model retains reasonable predictive power under realistic observational conditions.

In the left panel of Figure~\ref{fig:performance}, the HPD and TARP results for the cylindrical setup are shown as solid lines. Compared to the idealised spherical setup, the coverage curves deviate more from the reference line. However, both curves remain in reasonable agreement with perfect coverage, indicating that the predicted posteriors remain well-calibrated under more realistic observational conditions.
 
In the right panel of Figure~\ref{fig:performance}, the NPE-predicted cluster masses exhibit substantially less scatter around the true values compared to the \Msigma estimates computed from member galaxies identified by the ML membership identifier network. The NPE posterior standard deviation remains in good agreement with the empirical scatter, confirming consistent uncertainty calibration. The slight divergence between the NPE standard deviation and the empirical standard deviation near $10^{15}\,M_\odot$ is likely attributable to the limited number of clusters in this mass range within the test set. The predicted bias remains close to zero across the full mass range, again, indicating no significant systematic offset.
 
As expected, the overall mass inference performance for the cylindrical setup is worse than for the idealised spherical case, reflecting the impact of interloper contamination. Nevertheless, for clusters more massive than $10^{14.5}\mathrm{M_\odot}/h$, the performance approaches that of the spherical setup. This high-mass range scatter is approximately 25\% smaller than the $\sim 0.133$ dex value obtained in \cite{Ho2019} in the same high-mass regime, and is comparable to the performance suggested by \cite{Tominaga2026}. Since neither study reports scatter specifically for clusters above $10^{14.5}\mathrm{M_\odot}/h$, these comparisons should be regarded as approximate estimates based on their published figures. This suggests that our model can effectively correct for interloper-induced bias in the high-mass regime, even in the presence of significant interlopers.

At the low-mass end, we find larger scatter, both relative to higher-mass clusters in our sample and compared to the same mass range in \cite{Ho2019} and \cite{Tominaga2026}. We suspect that this difference is driven primarily by our inclusion of central galaxies from neighboring halos, as mentioned in Section~\ref{sec:cylinder_feature_construct}. In particular, for clusters with masses around $10^{14}\,\mathrm{M_\odot}/h$, interlopers account for approximately 84\% of the galaxies in the contaminated catalog. At these low-richness and high-contamination levels, the dynamical information carried by the true cluster population could be substantially diluted, making mass inference more challenging.

\section{Conclusion}
\label{sec:conclusion}
We present a simulation-based, fully probabilistic framework for galaxy cluster mass estimation that leverages the full projected phase-space information of cluster galaxies. By combining a permutation-invariant Deep Sets architecture with neural posterior estimation via normalizing flows, our approach moves beyond traditional summary statistics such as velocity dispersion and instead extracts higher-order dynamical information directly from the galaxy distribution.

A central feature of our method is the decomposition of the prediction into a baseline \Msigma\ estimate and a learned residual correction. This design isolates the information beyond equilibrium-based assumptions while preserving interpretability. Across both idealized and realistic setups, the model consistently improves upon the classical \Msigma\ relation, reducing scatter to lognormal residuals as low as $\sim 0.1$ dex in the idealized case and achieving comparable performance for massive clusters ($> 10^{14.5}\,M_\odot/h$) in the cylindrical setup.

Our framework produces full posterior distributions rather than point estimates. These posteriors are well calibrated, as demonstrated by HPD and TARP coverage tests and by the consistency between predicted uncertainties and empirical scatter. This enables a more complete characterization of uncertainty, including asymmetric and multimodal cases.

In realistic observational conditions, we incorporate a Deep Sets based interloper classifier that significantly outperforms the caustic method. While interloper contamination degrades performance, the full pipeline remains robust, particularly in the high-mass regime where it approaches the idealized limit. Importantly, retaining interloper information during inference provides additional environmental context that improves mass estimation.

Overall, our results show that simulation-based inference with set-based neural architectures offers a powerful and flexible framework for cluster mass estimation. By combining physical priors with expressive probabilistic models, this approach achieves improved accuracy and reliable uncertainty quantification, making it well suited for upcoming large-scale cosmological surveys.

\section*{Acknowledgments}

We thank Wooseok Kang for valuable discussions on the caustic method, including its implementation and limitations, which significantly improved our analysis. BYW thanks Hy Trac for helpful discussions during the early stages of this project.
M.H. is supported by the Simons Collaboration on ``Learning the Universe''. 
LT thanks Masahiro Takada for useful discussions.
LT is supported by JSPS under KAKENHI 24K22878 and 26K17136 and by the Royal Society under ICA\textbackslash R2\textbackslash 252140.
The Kavli IPMU is supported by World Premier International Research Center Initiative (WPI), MEXT, Japan.

\bibliographystyle{apsrev4-1}
\bibliography{main}

\appendix
\section{Interloper Identification Performance}
In this section, we show additional evaluation for the performance of the caustic method and our machine learning model for interloper identification.

\begin{figure*}[!htb]
    \centering
    \includegraphics[width=0.508\linewidth]{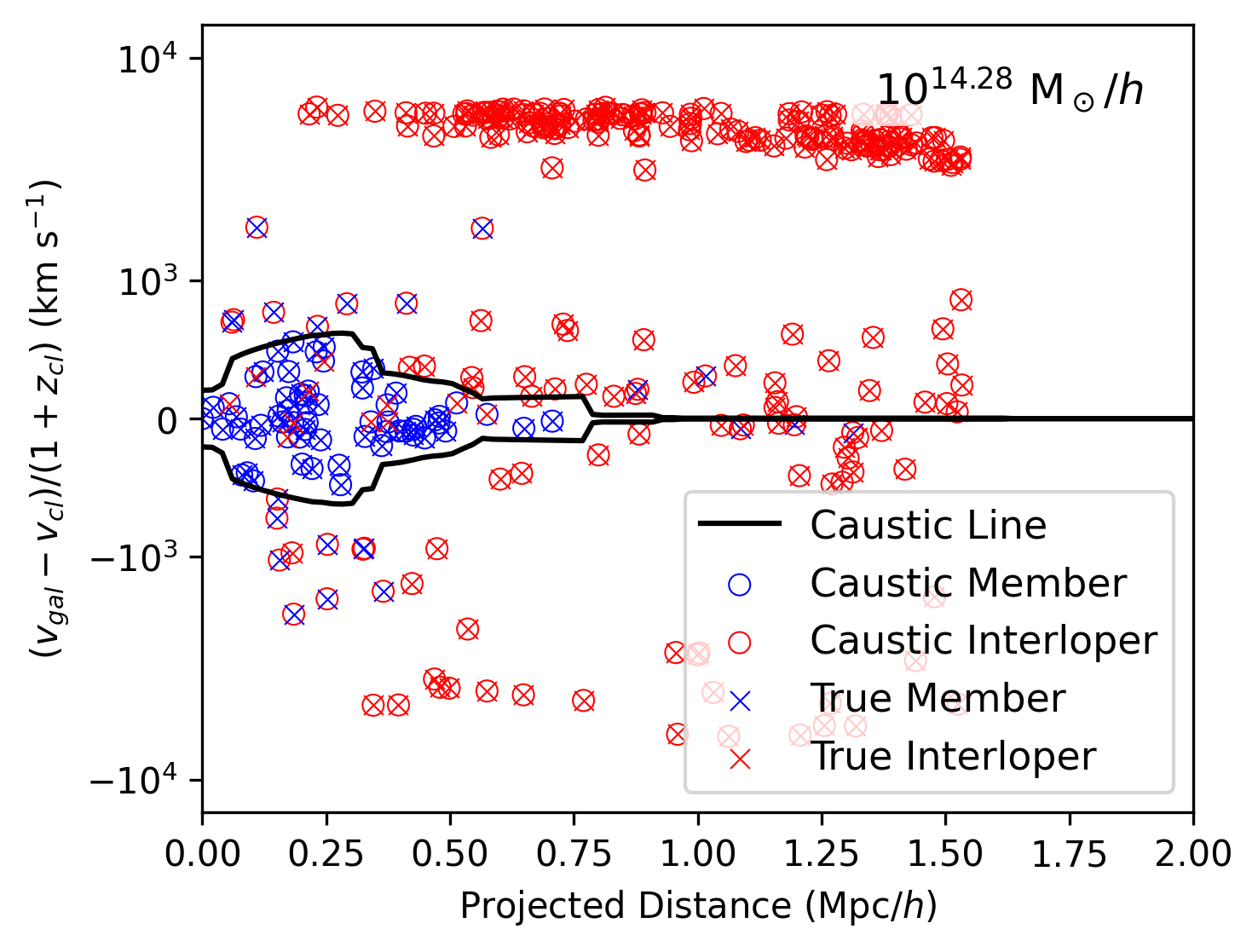}
    \includegraphics[width=0.482\linewidth]{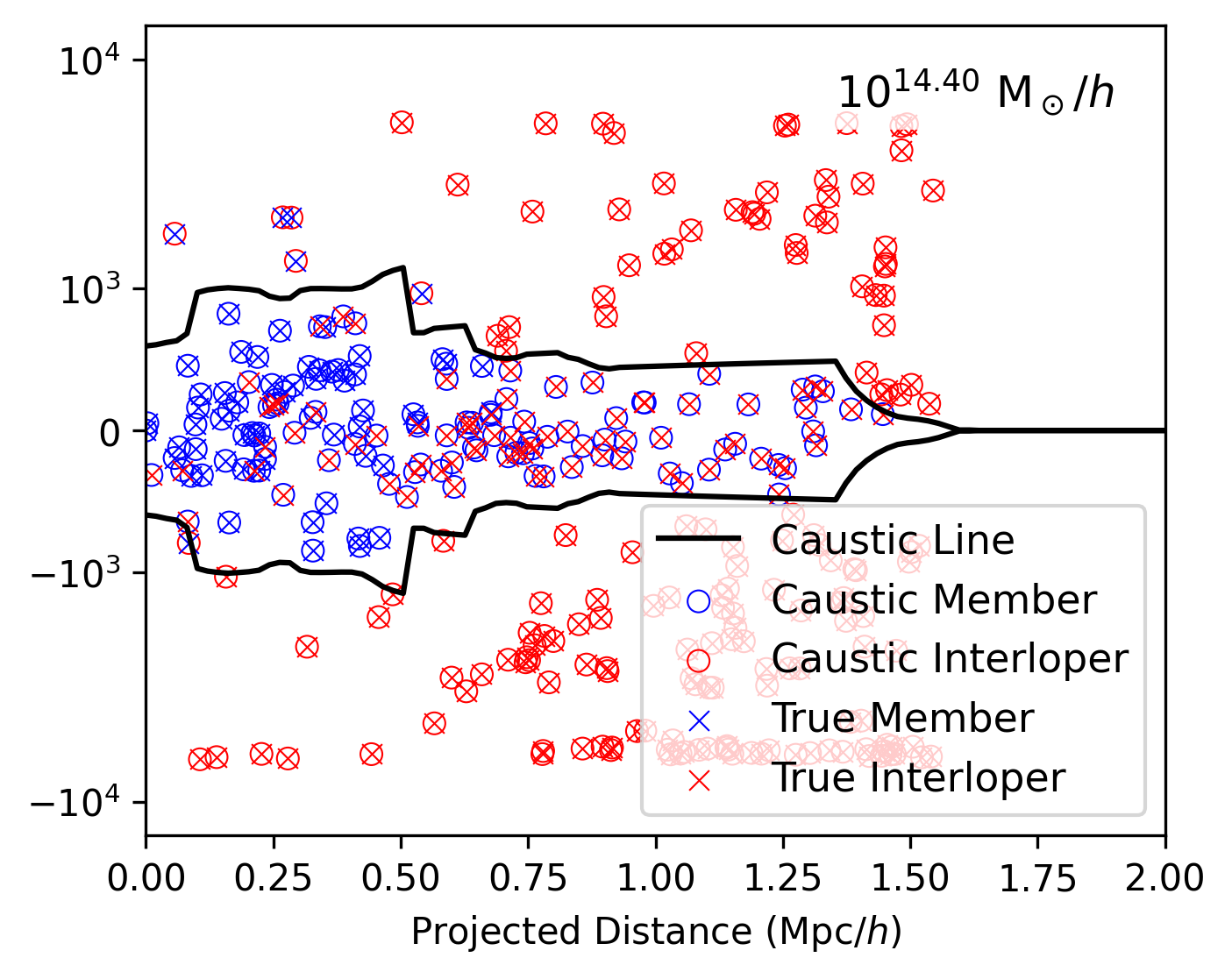}
    \caption{Comparison of interloper identification using the caustic method in projected phase space for two representative galaxy clusters. The horizontal axis shows the projected cluster-centric distance, while the vertical axis represents the line-of-sight velocity relative to the cluster center. The black curves denote the caustic boundaries used to separate members from interlopers. Blue circles indicate galaxies classified as members by the caustic method, and red circles indicate those classified as interlopers. Cross markers denote the ground truth from the simulation (blue for true members and red for true interlopers).}
    \label{fig:caustic-clusters}
\end{figure*}

Figure~\ref{fig:caustic-clusters} illustrates the performance of the caustic method implemented in \texttt{CausticSNU} in separating cluster members from interlopers in projected phase space for two representative systems. In both panels, the black curves trace the inferred caustic boundaries, which define the escape velocity profile and are used for membership classification. Blue circles denote galaxies identified as members by the caustic method, while red circles indicate those classified as interlopers. Cross markers represent the ground truth from the simulation.

We note that the projected distance used in the caustic analysis differs from the comoving projected distance $R_\perp$ defined above. Instead, galaxy positions are first transformed from projected polar coordinates into sky coordinates (RA, Dec). The angular separation of galaxies and cluster centers $\alpha_{\mathrm{caustic}}$ is then evaluated. The angular diameter distance at the cluster redshift $z_{\rm cl}$ is computed as $d_A(z_{\rm cl})$. The projected distance used in the caustic method is given by
\begin{equation}
R_{\mathrm{caustic}} = d_A(z_{\rm cl}) \sin\alpha_{\mathrm{caustic}}.
\end{equation}

The left panel of Figure \ref{fig:caustic-clusters} presents a relatively well-determined cluster, where the majority of true members are correctly enclosed within the caustic envelope. Only a small fraction of interlopers are incorrectly included, primarily at larger projected radii or near the caustic boundary. In contrast, the right panel demonstrates a more complex dynamical environment, where the separation between members and interlopers is less distinct. As a result, the caustic method both misclassifies true members as interlopers and includes a larger number of contaminants within the caustic boundary. This example highlights the limitations of analytical phase-space cuts in the presence of substructure or projection effects. This degradation in performance may also be exacerbated by the imposed stellar mass cut of $M_* > 10^{9.5}\,M_\odot/h$, which removes low-mass galaxies in the cluster outskirts. Since these galaxies provide important tracers of the phase-space structure at large radii, their absence could make the determination of the caustic boundary less robust.

\begin{figure*}[!htb]
    \centering
    \includegraphics[width=0.49\linewidth]{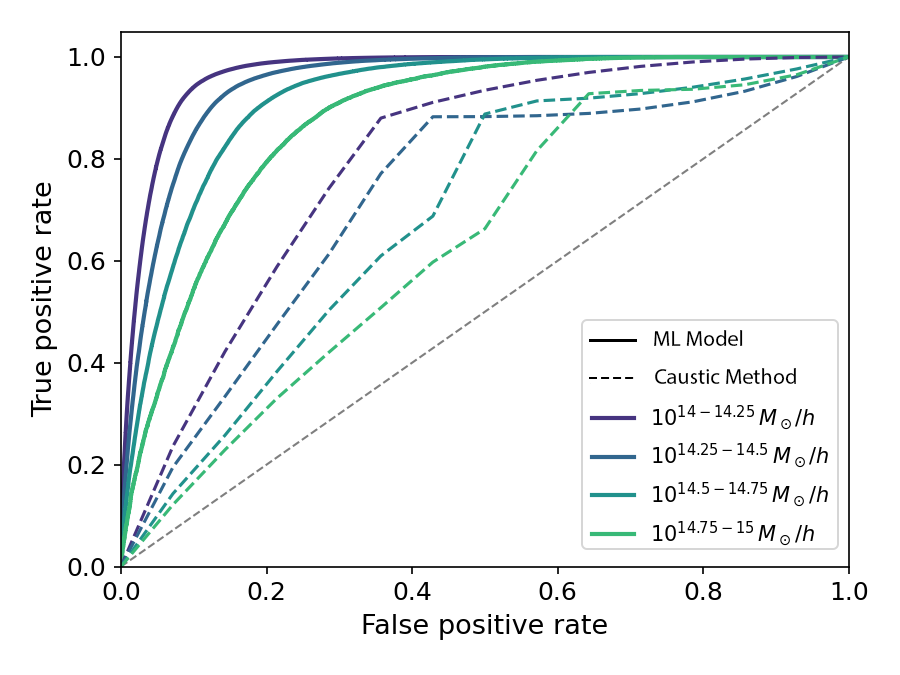}
    \includegraphics[width=0.49\linewidth]{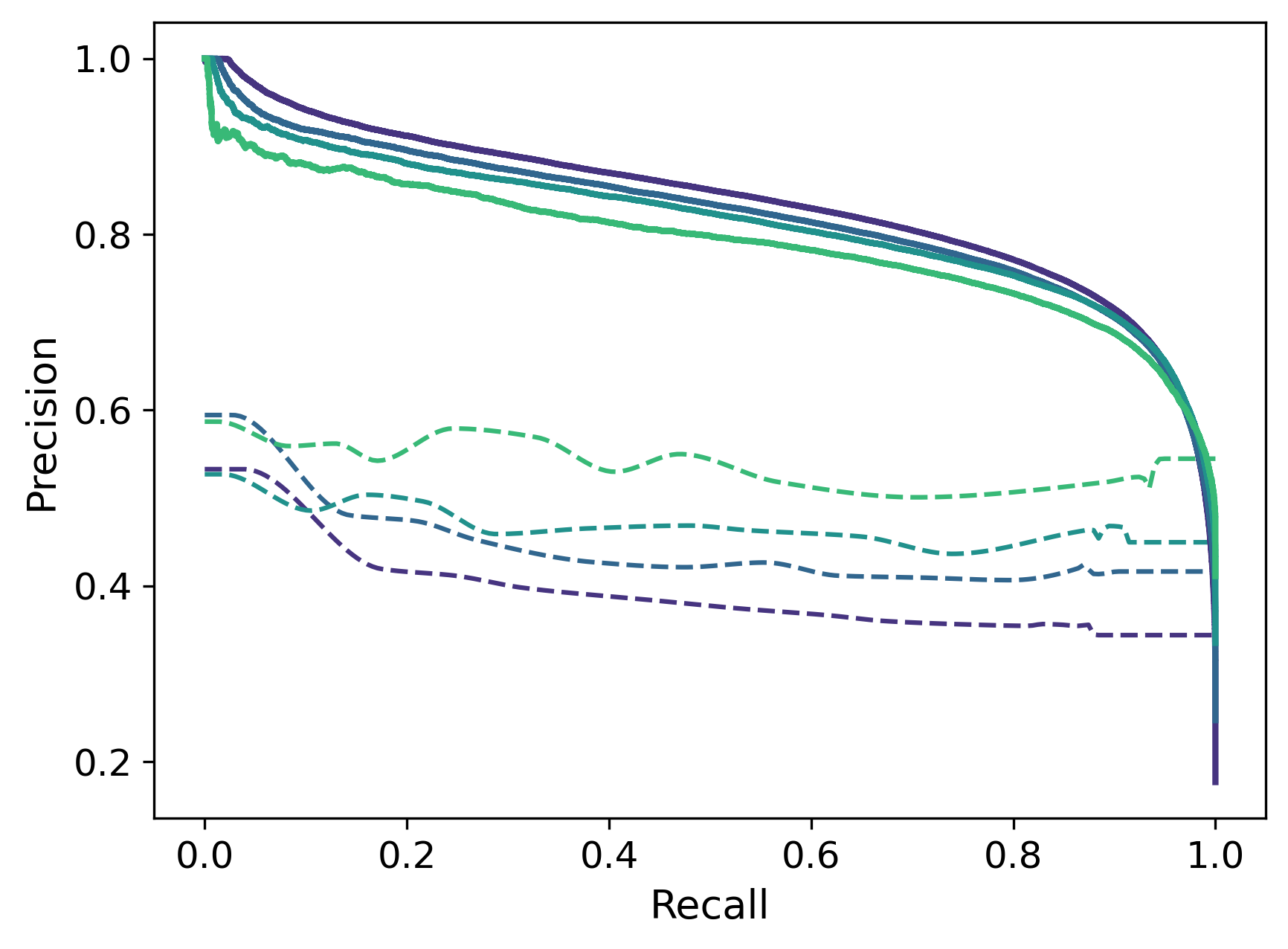}
    \caption{Performance comparison of the machine learning and caustic interloper identification methods across different cluster mass bins. Left panel: Receiver Operating Characteristic (ROC) curves showing the true positive rate as a function of the false positive rate. Right panel: Precision-Recall (PR) curves illustrating the trade-off between precision and recall. Solid lines correspond to the machine learning model, while dashed lines represent the performance of the caustic method. Each color denotes a different halo mass range, as indicated in the legend.}
    \label{fig:roc}
\end{figure*}

In figure \ref{fig:roc}, we use both the Receiver Operating Characteristic (ROC) and precision-recall (PR) diagnostics to compare the performance of the machine learning interloper identification model with the traditional caustic method across different cluster mass bins. Solid lines correspond to the machine learning model, while dashed lines represent the caustic method. For the machine learning model, the curves are generated by sweeping over the predicted membership probability threshold, while for the caustic method, the curves are obtained by varying the phase-space boundaries through the maximum velocity and maximum distance parameters in {\texttt{CausticSNUpy}}.

In the left panel, the ROC curves show that the machine learning model consistently outperforms the caustic method across all mass bins. The ML curves rise steeply toward the top-left corner, indicating high true positive rates at low false positive rates. In contrast, the caustic method exhibits significantly shallower curves, reflecting its limited ability to cleanly separate members from interlopers in projected phase space. The right panel presents the corresponding precision-recall curves. The machine learning model maintains high precision across a broad range of recall values, whereas the caustic method shows substantially lower precision, typically remaining below $\sim 0.6$ even at moderate recall. The ML model exhibits a more favorable trade-off, retaining significantly higher precision at comparable recall levels.

Overall, these results again demonstrate that the machine learning approach provides a substantial improvement over the caustic method in interloper identification for our contaminated cylindrical dataset. The gains are consistent across all mass bins.
\label{app:interloper_performance}

\label{lastpage}

\end{document}